\documentclass[aps,pre,twocolumn]{revtex4}
\usepackage{epsfig}   

\newcommand{\rhor}{\rho_{\scriptscriptstyle\rm R}}
\newcommand{\rhoc}{\rho_{\scriptscriptstyle\rm c}}
\newcommand{\kb}{k_{\scriptscriptstyle\rm B}}
\newcommand{\tw}{t_{\scriptscriptstyle\rm w}}

\begin{document}

\title{Slow dynamics under gravity: a nonlinear diffusion model} %

\author{Jeferson J. Arenzon}
\email{arenzon@if.ufrgs.br}
\affiliation{Instituto de F{\'\i}sica, Universidade 
         Federal do Rio Grande do Sul\\ 
         CP 15051, 91501-970 Porto Alegre RS, Brazil}

\author{Yan Levin}
\email{levin@if.ufrgs.br}
\affiliation{Instituto de F{\'\i}sica, Universidade 
         Federal do Rio Grande do Sul\\ 
         CP 15051, 91501-970 Porto Alegre RS, Brazil}

\author{Mauro Sellitto}
\email{sellitto@ictp.trieste.it}
\affiliation{Abdus Salam International Centre for Theoretical Physics \\
         Strada Costiera, 11, 34100 Trieste, Italy}

\date{\today}  

\begin{abstract}
        We present an analytical and numerical study of a nonlinear
	diffusion model which describes density relaxation of loosely
	packed particles under gravity and weak random (thermal)
	vibration, and compare the results with Monte Carlo
	simulations of a lattice gas under gravity.  The dynamical
	equation can be thought of as a local density functional
	theory for a class of lattice gases used to model slow
	relaxation of glassy and granular materials.  The theory
	predicts a jamming transition line between a low density fluid
	phase and a high density glassy regime, characterized by
	diverging relaxation time and logarithmic or power-law
	compaction according to the specific form of the diffusion
	coefficient.  In particular, we show that the model exhibits
	history dependent properties, such as quasi
	reversible-irreversible cycle and memory effects -- as observed
	in recent experiments, and dynamical heterogeneities.  
\end{abstract}


\maketitle


\section{Introduction}

The dynamics of granular matter has received considerable attention in
the past few years as it poses interesting problems from a theoretical
point of view, besides its relevance to industrial
applications~\cite{Edwards94,deGennes98,Kadanoff99,Duran00}. At high
density, excluded-volume interactions play a crucial role in the
formation of disordered, amorphous granular packings. In fact, the
analogy between slowly compacting granular materials and other
disordered systems like glasses has been early
recognized~\cite{Struik78,Edwards91}, and has motivated several
experiments in which slow compaction and history dependence have been
investigated in great
detail~\cite{KnFaLaJaNa95,NoKnPoJaNa97,NoKnBeJaNa98,ViLaMuJa00,DaGr01b}.

Granular and glassy systems share the important feature of having an
exponentially large number (in the system size) of different
mechanically stable packings.  Microscopically, this property can be
thought of as generated by geometric frustration or kinetic
constraints on the possible moves or positions of particles.  This in
turn leads, at high packing density, to a vanishing particle mobility
which is the distinctive macroscopic manifestation of slow relaxation
and jamming transition (dynamical arrest).

In general, two ingredients are responsible for the unusual behavior
of a granular material. First, collisions between the particles are
inelastic, and energy has to be constantly pumped into the system.
Second, at high packing density, the excluded volume, and the
associated cage effect, is very similar to the one observed in
structural glasses~\cite{WaHa96}.  A number of schematic lattice-gas
models~\cite{NiCoHe97,BrPrSa99,SeAr00,BaLo00,PhBi01,BeMe01,FuFaGaPeRo02}
has shown indeed that the main features of irreversible compaction do
not depend, to a certain extent, on the dissipation mechanism (which
is often assumed to be, for simplicity, of thermal nature), but can be
understood solely in terms of steric hindrance.  In this paper, we
shall be concerned precisely with this quasi-static flow regime, which
in the case of dense systems is the relevant one.  Even with this
simplifying assumption the detailed correspondence with mean-field,
mode-coupling approaches of glassy dynamics remains however
problematic because the presence of gravity leads to a non-trivial
dependence on the spatial variable for the basic observables. Other
complications may further arise from the presence of boundary
conditions.  At the present stage, coarse-grained approaches based on
real-space diffusion equations can therefore be very useful in the
theoretical interpretation of experimental results and to disentagle
the glassy features which are inherent to the compaction dynamics from
the ones which depend on the specific energy injection/dissipation
mechanism.

Some typical questions in slow granular dynamics that one is concerned
with are the:

\begin{itemize}

\item origin of the logarithmic compaction law;

\item scaling behavior of the aging dynamics;

\item reversible-irreversible cycle; 

\item memory phenomena;

\item and dynamical heterogeneities. 

\end{itemize}

We have recently addressed some of these issues by studying the
dynamics of a kinetic free-volume model for granular
media~\cite{SeAr00}, and proposing an analytical approach based on a
dynamical local density functional theory~\cite{LeArSe01}.  We have
precisely characterized the way compaction and aging depend on the
particle mobility of a homogeneous system.  Specifically, there are
two relaxation regimes, fast and slow, separated by a dynamic jamming
transition. The compaction law in the slow dynamic regime depends upon
the particle mobility: we find that logarithmic compaction and simple
aging are intimately related to the Vogel-Fulcher law, while power-law
compaction and superaging behavior occur in the presence of a
power-law vanishing mobility.  The objective of this paper is to
extend our previous work~\cite{SeAr00,LeArSe01}, and in particular to
address other history dependent phenomena, such as
reversible--irreversible
cycles~\cite{NoKnPoJaNa97,NoKnBeJaNa98,ViLaMuJa00}, memory
phenomena~\cite{JoTkMuJa00,NiDuPo00,BrPr01,BaLo01} and dynamical
heterogeneities~\cite{Sillescu99,Richert02}, which have recently
attracted some attention.

In the next section we shall introduce the nonlinear diffusion model
and the related lattice-gas which is thought to be its microscopic
realization.  The stationary density profiles, along with the jamming
transition will be presented in section~\ref{section.profile}.  In
section~\ref{section.compaction} the possible scenarios for the time
evolution of packing density are described.  Reversible-irreversible
cycles and memory phenomena will be discussed in
section~\ref{section.cycles}, while the scaling behavior of the aging
dynamics is addressed in section~\ref{section.aging}.
Section~\ref{section.het} discusses the dynamical heterogeneities in
presence of gravity and, finally, in section~\ref{section.conclusions}
the conclusions will be presented.

\section{Models}
\label{section.model}

\subsection{Nonlinear diffusion equation}

We assume that the dynamical evolution of the local particle density
$\rho(z,t)$ is governed by the continuity equation, 
	$$
	\frac{\partial \rho(z,t)}{\partial t}
	+\frac{\partial J(z,t)}{\partial z} = 0\;\; , 
	$$
with the particle current $J(z,t)$ given by the Fick's law,
	$$ 
	J(z,t)=-\Gamma(\rho) \frac{\partial
	\mu(z,t)}{\partial z} \;\;,
	$$ 
where $\Gamma(\rho)$ is the Onsager mobility and $ \mu(z,t)=
\frac{\delta F}{\delta \rho}$ is the local chemical potential.  The
only interaction between the grains we consider is the hard core
repulsion, for which the exact lattice Helmholtz free energy
functional is~\cite{Levin00}
\begin{equation}
\label{2}
	\beta F[\rho(z,t)]= \int_0^H \!\!dz\left[
	\gamma \, z \, \rho - S(\rho) \right] \,,
\label{eq.functional}
\end{equation}
where the entropy $S(\rho)$ is given by 
\begin{equation}
	S(\rho)= - \rho\ln \rho - (1-\rho)\ln(1-\rho) \,.
\label{eq.entropy}
\end{equation}
For highly packed hard-sphere systems, theoretical and experimental
studies suggest that the diffusion coefficient vanishes as a
power-law~\cite{GoSj92,ToOp95,MeMoWiMu98}. Hence we will assume that
the mobility $\Gamma(\rho)$ vanishes as
$$
	\Gamma(\rho)= 
	\Gamma_0 \, \rho \, \left(1-\frac{\rho}{\rhoc}\right)^\phi\;, 
$$ 
and remains zero for $\rho>\rhoc$.  Below we will also discuss another
possible functional form for the mobility which is commonly
encountered in systems of particles with aniso\-tropic shape
(e.g. rods).  Note that the above functional form of mobility has an
implicit dependence on the height, because the density profile is
typically inhomogeneous ($\rho=\rho(z)$) due to the driving force and
boundary conditions.  The use of a local density approximation for the
mobility will be justified by the comparison of theoretical
predictions with Monte Carlo simulations of a lattice gas model which
exhibits a vanishing diffusion coefficient at a threshold density
$\rhoc$~\cite{KoAn93}.  Substituting $\Gamma(\rho)$ into the
continuity equation we are led to
\begin{equation}
\label{eq.diffusion}
	\frac{\partial \rho}{\partial t}=\frac{\partial}{\partial z}
	\left\{ \left( 1-\frac{\rho}{\rhoc} \right)^\phi
	\left[\frac{1}{1-\rho}\frac{\partial \rho}{\partial z} +
	\gamma \, 
	\rho \right]\right\}\;,
\end{equation}
where the time is now measured in units of $1/\Gamma_0 \kb T$.  This
equation has to be completed by specifying the boundary conditions.
We will discuss two simple cases corresponding to open and closed
systems. In both situations one boundary condition requires the
vanishing of the current at the bottom layer $z=0$, $J(0,t)=0$ for any
time $t$. If the top layer $z=H$ is in contact with a particle
reservoir at density $\rhor$ (open system), the other boundary
condition reads $\rho(H,t)=\rhor$ for all $t$; while for a closed
system in which the total number of particles is kept constant, the
second boundary condition leads to a vanishing current also at $z=H$,
$J(H,t)=0$.  Although no closed analytical solutions of
Eq.~(\ref{eq.diffusion}) is found, it is possible to characterize its
asymptotic long time regime by an explicit calculation of density
relaxation and two-time mean-square displacement.

\subsection{Microscopic lattice gas}

\begin{figure}[h]
\epsfig{file=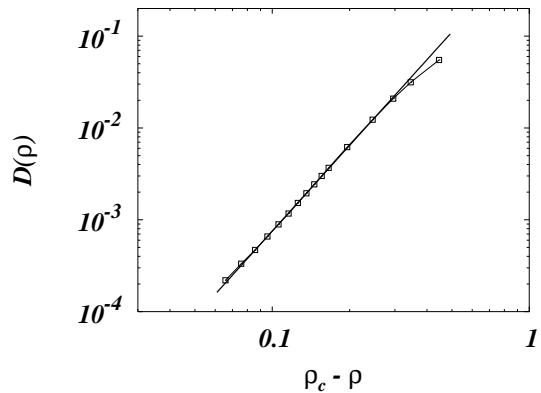,width=5.5cm, angle=270}
\caption{The diffusion coefficient of a BCC lattice ($L=32$) as a
function of the density. The full line is a power-law fit with
$\rhoc\simeq 0.84$ and $\phi=3.1$.}
\label{fig.bcc}
\end{figure}

The simplest microscopic realization of the nonlinear diffusion
equation one can imagine is provided by a lattice gas having a
vanishing diffusion coefficient above a certain threshold density (see
\cite{RiSo02} and references therein).  A paradigmatic example is the
kinetically constrained lattice gas model devised by Kob and
Andersen~\cite{KoAn93}.  The model was originally introduced with the
purpose to test the predictions of the mode-coupling theory for
supercooled liquids~\cite{GoSj92}.  The system consists of $N$
particles on a lattice with at most one particle per site and no other
static interactions between the particles, that is the Hamiltonian is
${\cal H}=0$.
%
%
The microscopic dynamic is as follows: at each time step a particle
and one of its neighboring sites are chosen at random; the particle
can move to the new site if this site is empty and if the particle has
less than $\nu$ nearest neighbors occupied {\it before} and {\it
after} the move.  This kinetic rule is time-reversible and the
detailed balance is satisfied.
At high densities, the dynamics slows down because the reduced
free-volume makes it harder for a particle to satisfy the dynamic
constraints. There exists a critical density $\rhoc$ above which the
particles are so interlocked that no macroscopic structural
rearrangement is possible and the mobility falls to zero as a power
law, $D(\rho)\sim (\rhoc-\rho)^{\phi}$.  For the simple cubic
lattice~\cite{KoAn93}, $\rhoc \simeq 0.88$, while for the body
centered cubic (BCC) lattice one gets $\rhoc \simeq 0.84$, see
Fig~\ref{fig.bcc}.  The critical density $\rhoc$ is therefore non
universal, depending both on the lattice structure and the particular
choice of the dynamical constraint parameter $\nu$.  In both cases,
however, the value of the exponent is consistent with $\phi \simeq
3.1$.  The universality of the exponent $\phi$ was recently suggested
by Imparato and Peliti, who studied the diffusion on a face centered
cubic lattice for several choices of $\nu$~\cite{ImPe00}.

Since much of dynamical properties of both structural glasses and
dense granular materials are dictated by steric constraints, we have
generalized~\cite{SeAr00} the Kob-Andersen model in a gravitational
field.  The Hamiltonian now is
\begin{eqnarray}
	\beta {\cal H} & = & \gamma \sum_i z_i n_i
\label{H}
\end{eqnarray}
where $n_i=0,1$ is the occupation variable of the $i$-th site whose
height is $z_i$, $\gamma= m g/\kb T$ is the inverse gravitational
length and $g$ is the constant gravitational field acting in the $-z$
direction.  Since we are interested in the slow compaction regime, the
energy dissipation due to inelastic collisions is ignored.  The
thermal energy of the grains is negligible and $T$ is neither the
physical temperature nor the ``granular temperature'' usually
associated with the average kinetic energy, but rather a function of
the externally imposed vibration intensity.  In other words, we assume
that the random diffusive motion of ``grains'' produced by the
mechanical vibrations of the box can be modeled as a thermal bath of
temperature $T$ ~\footnote{The effect of dissipation can be studied by
allowing for violation of detailed balance, as recently done for
example in Ref.~\cite{FuFaGaPeRo02}.}.  The particles satisfying the
kinetic constraints can move according to the Metropolis rule with
probability min$[1,x^{-\Delta h}]$, where $\Delta h = \pm 1$ is the
vertical displacement for the attempted elementary move and $x =
\exp(-\gamma)$ represents the ``vibration amplitude''.  Particles are
confined in a box closed at the bottom and with periodic boundary
condition in the horizontal direction. At the top, the box can be
either closed or in contact with a particle reservoir.  We set the
constraint threshold at $\nu=5$.  The Markov process generated by the
kinetic rules is irreducible on the full configuration
space~\cite{SeAr00}, the static properties of the model are those of a
lattice-gas of non-interacting particles in a gravitational field, and
these can be easily computed. For example, the mean occupation of each
level is:
\begin{equation}
	\rho(z)=\frac{1}{1+{\rm e}^{\gamma (z+\eta)}} \,,
\label{eq.profile}
\end{equation}
where the Lagrange multiplier $\eta$ is determined by the global
density or the chemical potential, according to the statistical
ensemble.

\section{Jamming transition and density profiles}
\label{section.profile}

\subsection{Open system}

\begin{figure}[h]
\epsfig{file=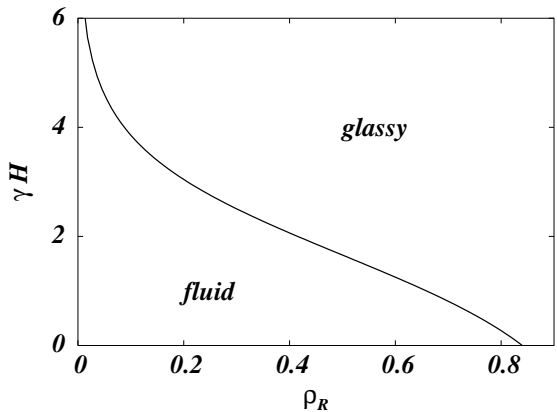,width=5.5cm, angle=270}
\caption{
	Jamming transition line $\gamma_c$ (multiplied by $H$) as a
	function of $\rhor$, from Eq.~\ref{eq.gammac}. This line
	separates regions of slow (glassy) and fast (fluid)
	relaxation.}
\label{fig.gammac}
\end{figure}

The stationary state of Eq.~(\ref{eq.diffusion}) is obtained when
$\partial \rho/\partial t=0$, which implies that $J(z,\infty)=0$ for
all $z$. Imposing the stationarity condition, depending on the value
of $\gamma$, two very distinct types of stationary profiles are found.
For high values of $\gamma$ the system is in equilibrium and the
profile is given by Eq.~\ref{eq.profile}.
As the vibration is lowered, a homogeneous region of constant density
$\rho(z,\infty)\equiv\rhoc$ develops below the height
$z_0$. The stationary solution of Eq.~\ref{eq.diffusion} becomes,
\begin{equation}
	\rho_{\infty}(z)=\left\{\begin{array}{l}
	\rhoc, \;\; z\leq z_0 \\ \mbox{} \\
	\displaystyle\frac{1}{1+\exp\left( \gamma z+\eta \right)}, \;\; z\geq z_0
	\end{array}
	\right.
\label{eq.rhoz}
\end{equation}
The values of $z_0$ and $\eta$ are obtained from the boundary
conditions.  If the top of the box is connected to a particle
reservoir then $\rho_{\infty}(H)=\rhor$. This and the continuity of
the density profile at $z=z_0$ lead to,
\begin{eqnarray}
        z_{\rm 0} &=& H + \frac{1}{\gamma} \ln \frac{\rhor (1-\rhoc) }{
	\rhoc (1-\rhor) } \\
	\eta &=&\ln \frac{1-\rhor}{\rhor} -\gamma H\;.
\end{eqnarray}
The jamming transition corresponds to the locus in the parameter space
$(\gamma,\, \rhor)$, at which the $z=0$ layer attains the critical
density so that $\rho(0,\infty)=\rhoc$.  This happens when
\begin{equation}
	\gamma_{\rm c}(\rhor) = \frac{1}{H}
	\ln  \frac{ \rhoc (1-\rhor) }{
	\rhor ( 1-\rhoc ) } \, .
\label{eq.gammac}
\end{equation}
When $\gamma > \gamma_{\rm c}(\rhor)$ the density at the bottom of the
box is close to the critical and dynamics becomes sluggish. On the
other hand, above the critical temperature, $\gamma \le \gamma_{\rm
c}(\rhor)$, all the layers have densities smaller than $\rhoc$ and the
system easily attains equilibrium. The critical line
Eq.~\ref{eq.gammac} is plotted, as a function of $\rhor$, in
Fig.~\ref{fig.gammac}. Notice that for the undriven case ($\gamma=0$)
the transition only occurs if $\rhor=\rhoc$.  It is important to
stress that for $\gamma>\gamma_{\rm c}(\rhor)$ the stationary profiles
Eq.~\ref{eq.rhoz} are not equivalent to the equilibrium ones since
they do not minimize the Helmholtz free energy functional
Eq.~\ref{eq.functional}.  This is so because the system is not able to
achieve, dynamically, densities higher than $\rhoc$.  In the zero
gravity case $\gamma=0$, the stationary profile is flat,
$\rho_{\infty}(z)=\rhor$.

\begin{figure}[h]
\epsfig{file=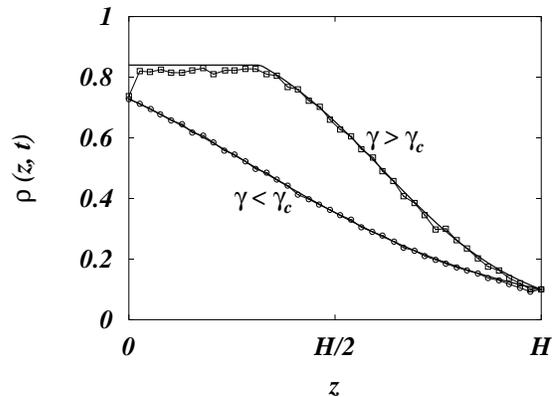,width=5.5cm, angle=270}
\caption{Stationary profiles above and below the critical line for an
	open system in contact with a particle reservoir at density
	$\rhor=0.1$.  The symbols are the densities obtained from the
	simulation on a BCC lattice and the solid lines are the
	theoretical results.  For $\gamma =0.072>\gamma_c$, squares,
	the longest time shown is $10^6$ MCS while for $\gamma
	=0.041<\gamma_c$, circles, the time is $10^5$ MCS.  Notice
	that the simulation profile for $\gamma >\gamma_c$ is not yet
	stationary.}
\label{profile}
\end{figure}

In Fig.~\ref{profile} we compare the stationary profiles with the ones
found in Monte Carlo simulation of the gravity-driven KA model on the
BCC lattice~\cite{SeAr00,LeArSe01} at large times.  
A very good agreement is obtained
with no adjustable parameters. As discussed in the next section,
the simulations were carried on
connecting the topmost layer of the system to a reservoir of
particles.

\subsection{Closed system}

\begin{figure}[h]
\epsfig{file=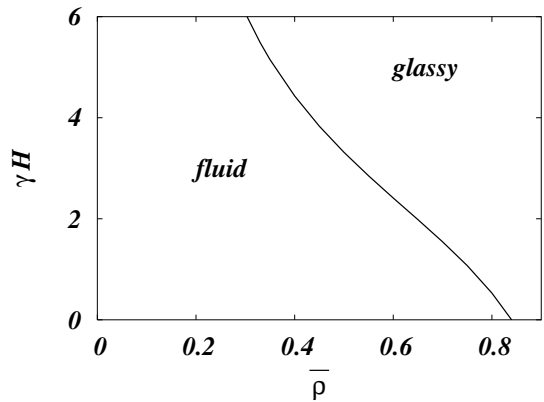,width=5.5cm, angle=270}
\caption{ 
	Jamming transition line, $\gamma_{\rm c}$ (multiplied by
	$H$), as a function of the average density $\overline{\rho}$.  
	The divergence
	near the origin scales as $\overline{\rho}^{\;-1}$, different from 
	the logarithmic divergence found for the system in contact with 
	a reservoir, Eq.~\ref{eq.gammac}. 
        Notice that in this case the density parameter
        is the average total density, $\overline{\rho}$, while in the open
        case it is the reservoir density $\rhor$.}
\label{fig.gamma_const}
\end{figure}

\begin{figure}[h]
\begin{center}
\epsfig{file=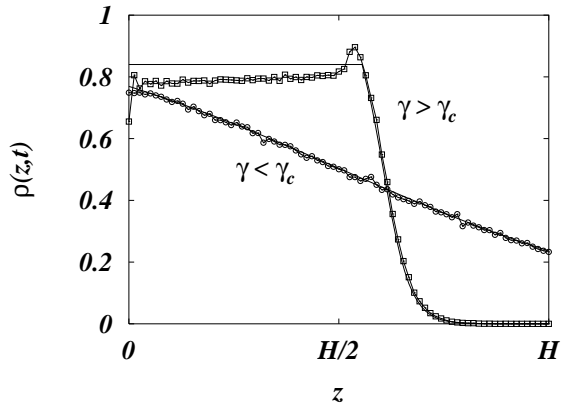,angle=270,width=8cm}
\end{center}
\caption{ 
	Examples of density profiles above and below the critical
	line for a system closed after its density achieved 
	$\overline{\rho}=0.5$.  
	The tapping amplitude in the glassy phase is $\gamma=0.4$,
	while in the fluid phase it is $\gamma=0.03$; system 
	height $H=80$. Notice that for $\gamma>\gamma_{\rm c}$ the
	profile at finite times ($10^6$ MCS in this case) is neither
	stationary nor flat.}
\label{fig.profile_closed}
\end{figure}

For a closed system of volume $V$ (the height $H$ times the
basis area) and a fixed number of particles
$N=\overline{\rho} V$ one proceeds in a way analogous to the previous section.
We find that $z_0$ and $\eta$ satisfy the coupled equations
\begin{eqnarray}
	z_{\rm 0} &=&
	\frac{1}{\gamma}\ln\frac{1-\rhoc}{\rhoc}-\frac{\eta}{\gamma}
	\label{eq.z0_const} \\
	\eta &=&\ln\frac{\mbox{e}^{-\gamma [\overline{\rho} H+z_0(1-\rhoc)]}
                 -\mbox{e}^{-\gamma H}}{
        1-\mbox{e}^{\gamma (\rhoc z_0-H\overline{\rho})}} \;.
	\label{eq.eta_const}
\end{eqnarray}
%
%
while the locus of the jamming transitions satisfies the implicit equation
\begin{equation}
	\gamma_{\rm c} = -\frac{1}{\overline{\rho} H} \ln \left(1-\rhoc+ \rhoc
	{\rm e}^{-\gamma_{\rm c} H} \right)
	\label{eq.gammac_const}
\end{equation}
%
This line is depicted in fig.~\ref{fig.gamma_const} as a function of
the average density $\overline{\rho}$. An example of a stationary profile for a
fixed number of particles is shown in fig.~\ref{fig.profile_closed}.
Simulations on bidimensional hard spheres also present a profile
compatible with an almost flat part plus an interface~\cite{ClRa91}.

We now turn to the behavior of the density profile found in the Monte
Carlo simulation.
We first let the system evolve at $x=0$ until a mechanically stable
configuration in which no particle can move down, is achieved. This
state is metastable because at $x=0$, there is a single ground state
with equilibrium bulk packing density $\rho = 1$.  Experimentally, it
has been verified that the initial density depends on the preparation
procedure~\cite{PoNiWe97}, in particular, on the box filling rate,
that is, the number of particles which fall per unit of time.  If
particles are poured one at a time, that is, they are individually
handled, then the falling trajectories are independent.  By doing so
in the gravity driven KA model, the kinetic constraints would always
be satisfied since the sites above the falling particle would be
empty, and the system would achieve a
fully compacted state.  
By pouring more than one particle at a time, their trajectories may
interfere, preventing the system from achieving the highest possible
density. This collective handling of particles may be implemented in
several ways. For example, in ref.~\cite{SeAr00}, in order to avoid
this highly compacted initial state, all the particles were poured
together, by placing them in the upper half of the box and letting
them fall randomly under the action of the gravitational field, until
a state where no particles can fall further is achieved.  In this case
the initial average packing density is $\rho_{\rm rlp} \simeq 0.707$,
roughly corresponding to a random loose packed state.
Once the system is prepared, the vibration at a fixed amplitude $x$ is
turned on. Another possibility, pointed out in the previous section is
to use a reservoir that may be left open forever or be closed after a
predetermined number of particles has entered the box.  This is also a
collective way of handling the particles, with the advantage that the
initial flux of particles can be tuned, being intermediate between all
particles falling at once and one at a time. Indeed, as the system is
less constrained than when all the particles are falling at once, it
is able to achieve a larger bulk density and the profile is closer to
the stationary one (see figs.~\ref{profile}
and~\ref{fig.profile_closed}).  Moreover, for the same reason, the
structured region at the bottom, discussed in \cite{SeAr00}, is either
absent or significantly less pronounced.

Near the top of the granular pile, a dense interfacial layer forms as
a result of an increased mobility due to the low density of particles
in the top most layers. This dense layer becomes more compact with
time hindering the underneath particles motion.
%
However, due to the (horizontal) roughness of the interface the effect
appears less pronounced than in the bottom region.
On the other hand, if the reservoir is kept open for all times, there
is no such sudden decrease in density, and even particles at the
interface are still quite constrained and no dense layer is observed
in the profile shown in the previous section (fig.~\ref{profile}).

\section{Compaction dynamics}
\label{section.compaction}

\subsection{Low density phase}

Above the jamming transition, $\gamma<\gamma_{\rm c}(\rhor)$,
the approach to equilibrium is exponentially fast, $\rho(z,t)\asymp
\rho_\infty(z)+g(z){\rm e}^{-t/\tau}$, as can be checked by
numerically solving Eq.~\ref{eq.diffusion}.


When the system is in permanent contact with a reservoir, the
characteristic time satisfies an {\it exact} scaling
equation~\cite{LeArSe01}
\begin{equation}
	\tau^{-1}=\frac{\pi^2}{4 H^2} {\cal F}(\gamma H;\rhor)\;,
\end{equation}  
where ${\cal F}(y;\rhor)$ is a scaling function. 
In the special case of zero gravity~\cite{PeSe98}, $\gamma=0$,
particles diffuse freely from the reservoir until a uniform density,
$\rho_\infty(z)=\rhor$, is established.  The characteristic time of
approach to equilibrium can be calculated explicitly~\cite{LeArSe01}
by linearizing Eq.~\ref{eq.diffusion}.  We find that the relaxation
time for $\gamma=0$ is
\begin{equation}
	\label{eq.tau0}
	\tau^{-1}=\frac{\pi^2}{4 H^2 (1-\rhor)}
	\left(1-\frac{\rhor}{\rhoc}\right)^{\phi} \,,
\end{equation}
or equivalently ${\cal F}(0;\rhor)=(1-\rhor/\rhoc)^{\phi}/(1-\rhor)$.
Eq.~\ref{eq.tau0} is in perfect agreement with the numerical
integration of Eq.~\ref{eq.diffusion}~\cite{LeArSe01}.  As expected,
the relaxation time diverges as $\rhor \rightarrow \rhoc$. The
exponent characterizing this divergence is $\phi$.

\begin{figure}[h]
\centerline{\epsfig{file=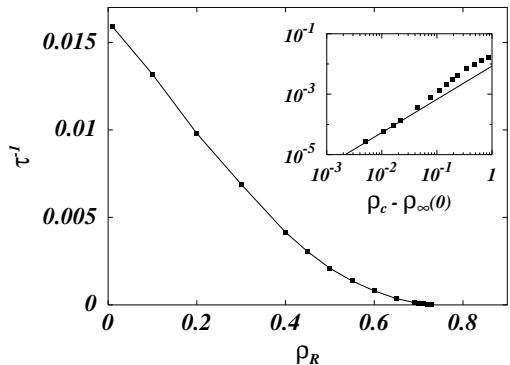,width=5cm,angle=270} }
\caption{
	Inverse relaxation time, $\tau^{-1}$, for $\gamma=0.1$ and $
	H=10$ with $\rhoc=0.88$ and $\phi=3.1$, corresponding to the
	Kob-Andersen model on a simple cubic lattice.  The points are
	the result of numerical integration of Eq.~\ref{eq.diffusion}.
	Inset shows the same data on the log scale. The characteristic
	time diverges with exponent $\phi-2$ as the jamming transition
	is approached.}
\label{fig.tau01}
\end{figure}

In the presence of a gravitational field we find, by numerical
integration of Eq.~\ref{eq.diffusion}, that as $\gamma \rightarrow
\gamma_{\rm c}(\rhor)$, the density of the first layer approaches
$\rhoc$, $\rho_\infty(0) \rightarrow \rhoc$, and ${\cal F} \sim
(\gamma_c-\gamma)^{\phi-2}$.  Thus, the relaxation time diverges with
exponent $\phi-2$, see fig.~{\ref{fig.tau01}}, implying that the
dynamics is faster than in the zero gravity case.  Comparing with
Eq.~\ref{eq.tau0}, we see that the jamming transitions in the
homogeneous and inhomogeneous systems, belong to distinct dynamic
universality classes.

\subsection{High density phase}

Below the jamming transition, $\gamma>\gamma_c(\rhor)$, the density of
the bottom layers, $z<z_{\rm 0}(\rhor)$, is close to the critical,
$\rho(z,t) \simeq \rhoc$, and the dynamics slows down.  To the lowest
order in $\Delta(z,t) \equiv 1-\rho(z,t)/\rhoc$,
Eq.~\ref{eq.diffusion} simplifies to,
\begin{equation}
	\label{4}
	\frac{\partial \Delta(z,t)}{\partial t}= -
	\gamma\frac{\partial \Delta^\phi}{\partial z} \,.
\end{equation}
To solve this non-linear equation we propose a scaling ansatz
$\Delta(z,t)=\Delta(z/t^\alpha)$.  Substituting into Eq.~\ref{4} we
see that this form is a solution if $\Delta(z/t^\alpha)$ is a power
law with $\alpha=1$:
\begin{equation}
\label{eq.deltaz}
	\Delta(z,t)=\left[\frac{z}{ \gamma \phi \;
	t}\right]^\frac{1}{\phi-1} \,.
\end{equation}
Notice, that in the absence of gravity, $\gamma=0$, density relaxation
is slower and characterized by a different dynamical exponent,
$\Delta(z,t) \sim t^{-1/\phi}$~\cite{PeSe98}.

Although in experiments one is usually interested in the bulk
properties, Eq.~\ref{eq.diffusion} can also shed some light on how the
upper layers $z>z_0$ compact.  Using the asymptotic solution,
Eq.~\ref{eq.deltaz}, in the definition of $J(z,t)$ we find that for
large times the particle current passing from the upper layers into
the bulk through $z=z_0$ is
\begin{equation}
	J(z_0,t) = \gamma \, \rho\left(\frac{z_0}{\gamma\phi \;
	t}\right)^{\frac{\phi}{\phi-1}} \;.
\end{equation}
Since the density of the upper layers is smaller than critical and
since we are only interested in the scaling behavior, it is sufficient
to study the linearized version of Eq.~\ref{eq.diffusion}
\begin{equation}
\label{eq.linear}
	\frac{\partial \rho}{\partial t}=\frac{\partial^2
	\rho}{\partial z^2} + \gamma \, \frac{\partial \rho}{\partial
	z} \,,
\end{equation}
with boundary condition: $\rho(H,t)=\rho_R$, and
$J(0,t)=J(z_0,t)$.  The temporal Laplace transform of
this linear equation can be easily solved yielding for the density
relaxation of upper layers the following expression:
\begin{equation}
	\rho(z,\infty)-\rho(z,t) \simeq \frac{J(z_0,t)}{\gamma} \left[
	{\rm e}^{\gamma(H-z)}-1\right] \sim t^{-\frac{\phi}{\phi-1}}\;
	\label{eq.delta}
\end{equation}
Remarkably, the time relaxation above and below $z_0$ are both slow
and follow a power law with different exponent. As expected, the
dynamics in the upper layers is faster than in the lower ones, and its
contribution to the relaxation function at long times becomes
negligible since $\phi$ is usually larger than one.

\begin{figure}[h]
\begin{center}
\epsfig{file=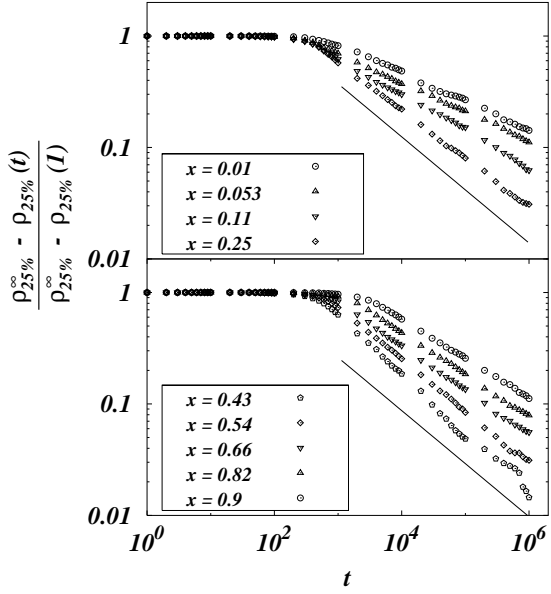,angle=270,width=9cm}
\end{center}
\caption{ Power law fit (normalized to unity), for large times, of
	compaction data.  The fit parameters are all dependent on $x$,
	although a unique exponent for the large values of $x$ is also
	consistent. The full straight line $t^{-1/(\phi-1)}$ represent
	the long-time prediction of the nonlinear diffusion equation.}
\label{fig.powerlaw}
\end{figure}

The asymptotic solution, Eq.~\ref{eq.deltaz}, is in partial agreement
with the lattice-gas Monte Carlo simulation data,
Fig.~\ref{fig.powerlaw}.  The same numerical data were previously
fitted with a four parameter logarithm law
\begin{equation}
	\rho(t) = \rho_{\infty} - \frac{\Delta\rho_{\infty}}{1 + B
	\ln\left( 1 + t/\tau \right)} \,,
	\label{eq.log}
\end{equation}
where $\rho_{\infty}$ is the asymptotic packing density and $B$,
$\Delta\rho_{\infty}$ and $\tau$ are adjustable parameters which also
depend on $x$.  The above function, first used in
Ref.~\cite{KnFaLaJaNa95}, gives a reasonable fit in the whole time
window accessible to experiments; however, one can check that the long
time behavior is also compatible with a power law relaxation.
Interestingly enough, something similar happens here, confirming that
a limited time-window may not allow to distinguish among several
regimes of slow relaxation~\cite{Head00}. One can also notice, from
fig.~\ref{fig.powerlaw} that for high values of $x$, all curves are
compatible with the same exponent, while for small values, the
exponents seems to be different. This may be however another artifact
of the very slow relaxation for small $x$, which prevents the system
from attaining the asymptotic regime.
 
\begin{figure}[h]
\begin{center}
\epsfig{file=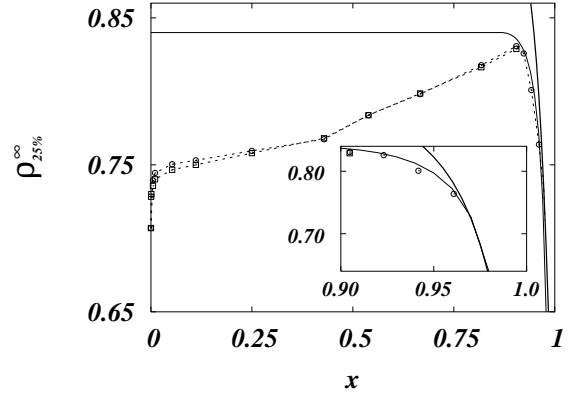,angle=270,width=8cm}
\end{center}
\caption{ 	
	Square and circle symbols are the asymptotic packing densities
	obtained within the Monte Carlo simulation with fixed number
	of particles. The two refer to the extrapolations using a
	power law fit and the logarithmic form, Eq.~\ref{eq.log}. The
	equilibrium density is the thick solid line while the thin
	line is the bulk density evaluated from the asymptotic limit
	of Eq.~\ref{eq.rhoz}, \ref{eq.z0_const} and~\ref{eq.eta_const}. 
	The dynamical jamming transition
	is located where all curves meet, near $x_c\simeq
	0.979$. Notice that, from the theory, up to the region near
	the maximum we have $z_0>H/4$ and the packing density is $\rhoc$,
	while above it the point $z_0$ penetrates in the defined bulk
	region and the density deviates from $\rhoc$.  Inset: region
	near the transition. Notice that the simulation data agree
        well with the result from the diffusion equation, while below
	$x_c$ both start to deviate from the equilibrium curve.}
	\label{fig.rho_inf_x_1}
\end{figure}

Further, we find that independently of the specific functional form of
the fit, logarithmic or power law, the asymptotic packing density
$\rho^{\infty}_{25\%}$ is quite the same for the simulation data,
Fig.~\ref{fig.rho_inf_x_1}.  It turns out to be a non monotonic
function of vibration $x$ and displays an optimal value for the
asymptotic compaction, as can be seen in Fig.~\ref{fig.rho_inf_x_1}.
%
%
This maximum is achieved for rather high vibration because an initial
decompaction increases the free volume available to particles making
it easier to satisfy the kinetic constraints and their local
arrangements.  
It is clear that the specific location of the maximum depends on the
portion chosen to measure the packing density ($H/4$ in this case)
however it does not affect the form of the plot.

%
%
%

\begin{figure}[h]
\begin{center}
\epsfig{file=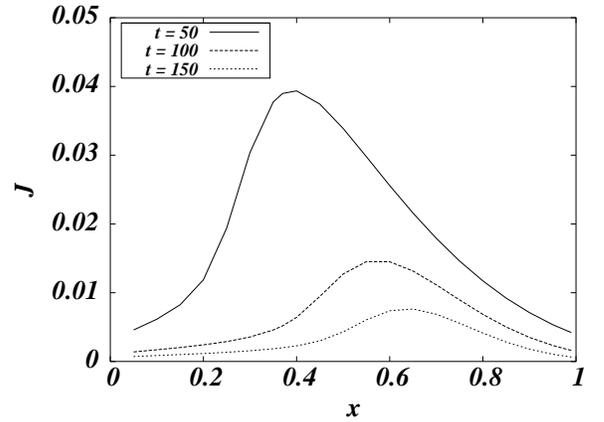,angle=270,width=8cm}
\end{center}
\caption{
	Local current $J(z,t)$ for several times $t$ and $z=H/2$,
	as a function of the vibration strength $x$.}
	\label{fig.nr}
\end{figure}

Finally, in order to stress the counterintuitive nature of compaction
dynamics it is worth to point out its relationship with the so-called
negative resistance phenomena. These are usually observed in a
non-equilibrium stationary state, where an increasing driving force
leads to a decreasing system response (usually a particle current).
During irreversible compaction, which is non-stationary, something
similar happens: indeed, when vibration (the driving force) increases
the {\it a priori} probability that a particle moves upward is larger;
it actually turns out that the system compacts, i.e. the average
direction of particle flow (the response) is preferentially downward
and non-monotonic.  This is illustrated in Fig.~\ref{fig.nr} where we
plot the local current $J(z,t)$ at different times $t$, as a function
of the vibration strength $x$. One observes that at any given time,
there is an optimum value of $x$ at which the current get its
maximum. At increasing time the maximum moves towards higher values of
$x$.  Qualitatively similar results are obtained for different $z$.

%
%

\subsection{Logarithmic compaction and Vogel-Fulcher law}

Up to now our discussion has been motivated by dynamics which are
characterized by a power-law vanishing mobility.  However, there are
various systems for which the mobility vanishes according to the
Vogel-Fulcher law~\cite{DaGr01b},
\begin{equation}
	\Gamma(\rho)= \Gamma_0 \, \rho \, \exp \left( 
	\frac{a\rhoc}{\rho-\rhoc} \right)\;.
	\label{eq.vogel}
\end{equation}
Boutreux and de Gennes~\cite{BoGe97} have argued that logarithmic
compaction is intimately related to this Vogel-Fulcher law which in
turns derive from a Poisson distribution of voids in the systems.
A similar conclusion can also be drawn from our
approach.  In this case the time evolution of the local particle
density is governed by the equation
\begin{equation}
\label{eq.diffusion_VF}
	\frac{\partial \rho}{\partial t}=\frac{\partial}{\partial z}
	\left\{ \exp \left( \frac{a\rhoc}{\rho-\rhoc} \right)
	\left[\frac{1}{1-\rho}\frac{\partial \rho}{\partial z} +
	\gamma \, 
	\rho \right]\right\}\;.
\end{equation}
Again, proceeding as in the previous section, to the lowest order in
$\Delta(z,t) \equiv 1-\rho(z,t)/\rhoc$, one finds that for the bottom
layers ($z<z_0$)
\begin{equation}
	\label{eq.vogel.asy}
	\frac{\partial \Delta(z,t)}{\partial t}= - \gamma
	\frac{\partial}{\partial z} \exp\left[ -\frac{a}{\Delta(z,t)}
	\right] \,.
\end{equation}
As before we attempt a scaling ansatz.  For very large times we find
\begin{equation}
	\Delta(z,t)= \frac{a}{ \ln (t/z)} \,,
\end{equation}
which is in agreement with the full numerical solution of
Eq.~\ref{eq.diffusion}.  This relaxation is similar to the logarithmic
compaction law, Eq.~\ref{eq.log} used to fit experimental data by
Chicago group~\cite{KnFaLaJaNa95}.

For layers above $z_0$, the time evolution is still slow but follows a
power law. Proceeding in the same way used to obtain Eq.~\ref{eq.delta}
we find
\begin{equation}
	\rho(z,\infty)-\rho(z,t) \sim t^{-1}\;.
\end{equation}
Although the bulk dynamics is logarithmically slow, the compaction of
upper layers is governed by a power law with exponent $-1$.

\section{History dependence}
\label{section.cycles}

It is well known that dynamical effects in slow relaxing systems
depend sensitively on the history of the sample after a quench in the
high density (or low temperature) phase. These phenomena have been
extensively studied by means of several experimental protocols.


In the previous section we focused on the simplest situation in which
the sample is prepared in a random loose packed state and then the
vibration is turned on and kept fixed to a given value of amplitude
during the measurement.

In order to test the system response, here we consider the effects of
cyclic changes in the vibration amplitude, either continuously or
suddenly, to a different value which corresponds to a state with a
lower or higher asymptotic packing density.

\subsection{`Reversible'-irreversible cycles}

Experiments on glass beads~\cite{NoKnPoJaNa97,NoKnBeJaNa98} have shown
that under a cyclic variation of the vibration a system prepared in a
random loose packed state first presents a branch during which the
density increases as a function of the vibration (until high
vibrations are attained and decompaction starts). This branch is
irreversible, meaning that when the vibration is decreased at the same
rate, the system does not trace back the earlier evolution, but rather
its density keeps growing as the vibration decreases.  Experimentally
this second branch appears to be reversible, that is, the system seems
to reach a stationary state in which any further cyclic variation of
the vibration keeps the system always on the second branch. Along this
branch the packing density is a decreasing function of the vibration
amplitude (contrary to what happens in the irreversible compaction
regime).

\begin{figure}[ht]
\begin{center}
\epsfig{file=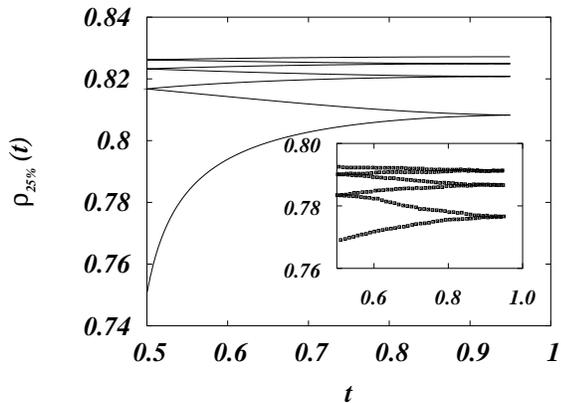,angle=270,width=8cm}
\end{center}
\caption{ Bulk density as a function of time for a cycling variation
	of the driving rate within the nonlinear diffusion equation
	(Eq.~\ref{eq.diffusion}). Here the waiting time is 500 steps
	and the reservoir is again at $\rhor=0.1$.  Each branch
	evolves from $5\times 10^3$ steps from the minimum $x$ value
	of 0.5 to the maximum one of $0.95$.  Inset: the same
	experiment with the gravity driven KA lattice-gas model. The
	parameters are the same as before, only that the waiting time
	is 5000 MCS and each branch has a total duration of $2\times
	10^4$ MCS.  In both cases, the distance between the branches
	decreases as the number of cycles increases.  }
\label{fig.cycle}
\end{figure}

\begin{figure}[ht]
\begin{center}
\epsfig{file=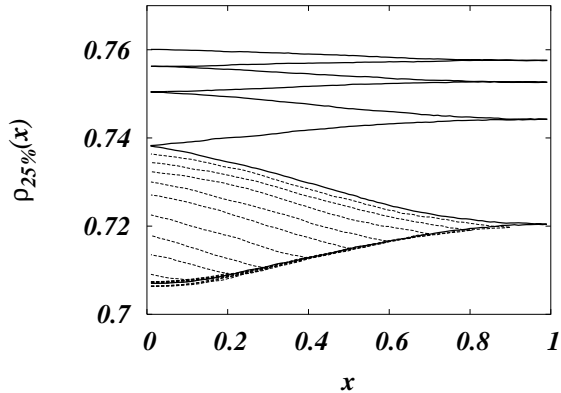,angle=270,width=8cm}
\end{center}
\caption{The same as the inset of fig.~\ref{fig.cycle} only that
$x$ is reversed at different maximum values. Notice that
in this case, different densities are attained.}
\label{fig.cycle2}
\end{figure}

Applying the same protocol to both the diffusion equation and the
lattice-gas we find that the reversible branch only appears for
extremely slow driving rates. Similar results have been observed
experimentally during the compaction of anisotropic granular materials
like rods~\cite{ViLaMuJa00} and shearing induced
compaction~\cite{NiDuPo00}. The system presents a succession of
irreversible branches which get closer and closer as the number of
cycles increases (see Figs.~\ref{fig.cycle} and the corresponding
inset).  The slower the vibration change rate is, the smaller is the
separation between the branches. In this way, for real systems the
distance between the irreversible branches can become of the order of
the measurement error, appearing as if there was just one reversible
branch. It must be noticed that after cycling for a certain number
of times, the density hardly changes from the lowest to the highest
values of $x$, presenting a rather flat behavior. This can be
explained using the asymptotic packing density,
fig~\ref{fig.rho_inf_x_1}: unless the vibration is too high, the
asymptotic bulk profile is flat. Even if the system enters in the
high-$x$ region, the flatness of the density will depends on the
vibration change rate. Finally, we mention that for some choices of the
parameters small hysteresis~\cite{BrPrSa00a} loops also appear in the 
region of high
vibration; the area of the loops is a function of the
vibration rate.  We also expect that by using the Vogel-Fulcher
mobility the cycles present a very similar behavior. Moreover,
for high vibrations, dissipation effects may play a role. It would be
interesting to experimentally study particles with different friction
properties to check to what extent the cycle properties depend on
these. In fig.~\ref{fig.cycle2} we illustrate the effect of
changing the turning point, that is, the maximum attained value of $x$
before starting to decrease it. We notice that the density follows
almost parallel paths, only that the maximum attained density is
bigger the higher is the turning point.


\subsection{`Memory' effects}

\begin{figure}[ht]
\begin{center}
\epsfig{file=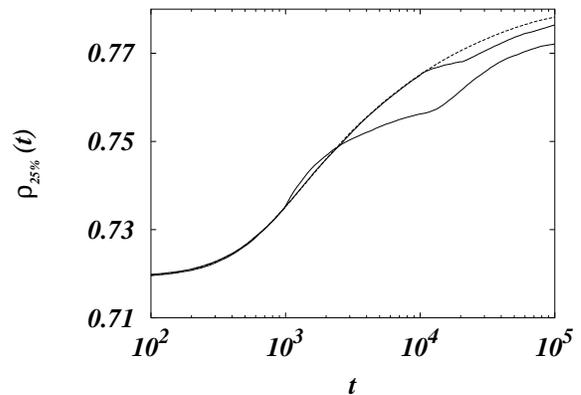,angle=270,width=8cm}
\end{center}
\caption{Short-term memory experiment performed with the gravity
driven KA model. The thick line shows 
the unperturbed evolution.
The system is perturbed at $t_w=10^3$ and $10^4$ after
evolving with $x=0.6$ (solid normal lines). 
In the interval $t_w<t<t_w+\Delta t$,
with $\Delta t=10^4$, the vibration changes to 0.3. Notice that
in the case where the system is perturbed earlier, the
system first increases its density (the short-term memory
effect as described in the text), resuming its expected
behavior after some time. In the case of a later perturbation,
this first regime is not noticeable.}
\label{fig.short_term_2}
\end{figure}

Another possible experiment devised to explore hystory dependence
consists of measuring the short-time response of an aged system to an
abrupt perturbation in $x$~\cite{JoTkMuJa00,NiDuPo00}.  After evolving
the system for a certain $t_w$ at a fixed $x$, the vibration amplitude
is shifted by $\Delta x$ until the time $t_w+\Delta t$ and then
returned to $x$.  As an example, in Fig.~\ref{fig.short_term_2}, we
show the curves for perturbations applied to systems with different
age, $t_w=10^3$ and $t_w=10^4$. In both cases, $\Delta x<0$, and the
compaction rate increases while one would expect, from the long term
behavior of the system, a slower relaxation for a smaller value of
$x$. For the older system, $t_w=10^4$, the first regime is hardly
visible because the system is stiffer.  These results, are consistent
with earlier experimental~\cite{JoTkMuJa00} and
theoretical~\cite{Head00,BrPr01,BaLo01} works.  For $\Delta x>0$ (not
shown), although one would expect a faster compaction, for short times
soon after the perturbation the system decompacts.  The same behavior
is also found for the diffusion equation when the system is in contact
with the particle reservoir. For large times after turning the
perturbation off, the system resumes its normal behavior. Thus,
because of the transient nature of the response, it appears as a
short-term memory. Moreover, after the initial anomalous behavior, the
perturbed and non perturbed curve crosses and the system start to
evolve with the rate expected from long term behavior.


\begin{figure}[ht]
\begin{center}
\epsfig{file=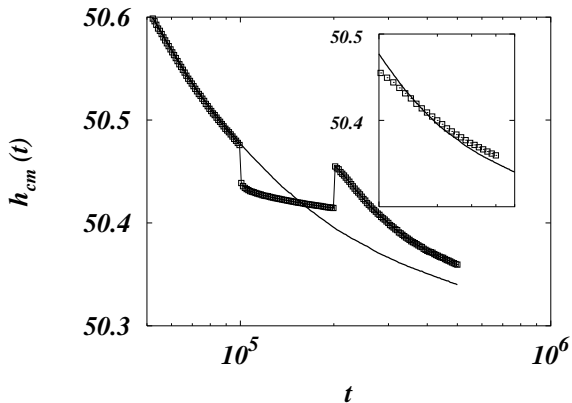,angle=270,width=8cm}
\end{center}
\caption{Long-term memory experiment. The evolution of the height
of the center of mass (or, equivalently, the potential energy) is
plotted as a function of time. Notice that if the system has
evolved enough time ($t_w=10^5$ in this case) the further
evolution after the perturbation is quite small and the system
returns to its previous state when the perturbation is turned
off. In the inset, the perturbed data for $t>t_w$ are shifted,
$t\to  t-10^5$ showing that they collapses on the unperturbed
data. Averages are over 300 samples.}
\label{fig.long_term}
\end{figure}

Besides short-term memory effects, long-term memory is also
present. To see this, we repeat the above experiments with
the difference that the perturbation is applied at a much
larger $t_w$ ($10^5$ in this case). Moreover, the perturbation is kept on 
also for a longer period. After being turned off, the system vibration
is returned to its previous values. As can be seen in
figure~\ref{fig.long_term} the
system evolves much less during the time interval $\Delta t$ when
it is perturbed: the $h_{cm}$
roughly follows a plateau. This is expected since the relaxation rate
decreases for smaller values of $x$. Moreover, when turning the perturbation
off, the system returns to a point very close to the one where
it was before being perturbed. This is more clear in 
the inset of the same figure, where the perturbed data
for $t>t_w+\Delta t$ are shifted by $\Delta t$ and seems, as
a first approximation,
to collapse on the unperturbed curve, showing that the
system keeps, to some extent, memory of its state before the perturbation
even if being perturbed for a long time.


As is common is systems in a non-stationary state, the response of the
system to a perturbation in the vibration clearly depends on its
age. The larger is $t_w$, the greater is the compaction achieved by
the bulk and less responsive the system becomes. For small $t_w$, the
bulk is still very sensitive to the perturbations, the amount of free
volume is considerable and as soon as the vibration is lowered, the
particles get closer and the density increases very fast, even faster
than one would expect from the knowledge of the asymptotic
behavior. By increasing, instead of decreasing, the vibration, the
opposite behavior is observed. Thus, short-term memory is related to
perturbations at early times.  However, as $t_w$ increases, the amount
of free volume in the bulk decreases and most of the instantaneous
response, when perturbed, comes from the interface, that has a fast
dynamics. After this strong response due to the interface (that is
only seem in global measures like the height of the center of mass),
the system continues at a much smaller pace, corresponding to the
expected evolution at the new vibration value. Thus, in the interval
$\Delta t$ while the perturbation is on, the amount of change is due
to the bulk evolution and is smaller the larger is $t_w$ (for example,
for the waiting times that we probed, from $10^3$ until $10^5$, the
bulk always evolved). As a conclusion, ``long-term memory'' effects
can only be seen as an approximation, with a very simple explanation
as also pointed out in \cite{BaLo01}.

As remarked in Refs.~\cite{Bouchaud02} and \cite{Kovacs63} (in the
context of glasses), a rate equation for a single macroscopic variable
(e.g., the free volume) would not be able to account for the
complexities of the memory effects. Nevertheless, it must be
emphasized that the nonlinear diffusion equation studied here is
explicitly dependent on the spatial dimension, and is this
heterogeneous profile that encodes the additional information
responsible for the effects discussed here. However, more complex
memory effects are observed in systems without gravity like glasses
and spin-glasses, and is unlikely that this equation would be able to
account for them, since probably they involve effects due to the
interaction.
In order to capture the complexity of these effects, this equation
would have to be properly generalized.

\section{Physical aging under gravity}
\label{section.aging}

We now turn to the discussion of physical aging phenomena under
gravity as they appear in the two-time scaling behavior after 
a sudden quench into the high packing density phase.

\subsection{Mean-square displacement}

In our approach the aging effects are best studied by considering the
two-times mean square displacement of particles, $B(t,t_{\rm w})$, which
for the 3D lattice-gas is defined as
\begin{equation}
	B(t,\tw)=\frac{1}{3N}\sum_{a=1}^{3} \sum_{k=1}^{N} 
  	\left\langle \left[ r_k^a(t+\tw)-r_k^a(\tw) \right]^2 
  	\right\rangle \,,
\end{equation}
where $r_k^a(t)$ are the coordinates ($a=1,2,3$) of the particles $k$
at times $t$.  In the continuous 1D diffusion model, if $t$ is
sufficiently larger than $t_{\rm w}$, the mean-square displacement at
height $z$ can be written as
\begin{equation}
	\label{5_1}
	B_z(t,t_{\rm w})= \int_{t_{\rm w}}^t ds \, \Gamma \left[
	\rho(z,s) \right] \,.
\end{equation}
It is worth to recall that in zero gravity one finds a simple
aging~\cite{PeSe98}.
For a power-law diffusion and for the lower layers $z<z_{\rm
0}(\rhor)$ we find, to leading order in $t$ and $t_{\rm w}$, a
two-time scaling of the form
\begin{equation}
	\label{5_2}
	B_z(t,t_{\rm w}) \sim t_{\rm w}^{1-\mu} - t^{1-\mu} \,,
	\label{B_a}
\end{equation}
with an exponent $\mu = \phi/(\phi-1)$.  Since, usually $\phi>1$, this
corresponds to a super-aging regime, $\mu > 1$.  This means that the
effective structural relaxation time grows as $t_{\rm w}^{\mu}$, what
is faster than the age of the system, $t_{\rm w}$. It also means that
there is a microscopic time scale that starts to become relevant,
differently from what happens with simple aging.  For the upper
layers, $z > z_{\rm 0}(\rhor)$ one obtains the same two-time scaling
behavior but this time the exponent is $\mu=\phi^2/(\phi-1)$.  
However, for closed systems at not so large vibrations, 
the contribution from the layers above $z_0$ is small since most
of the particles are at $z<z_0$.

\begin{figure}[h]
\begin{center}
\epsfig{file=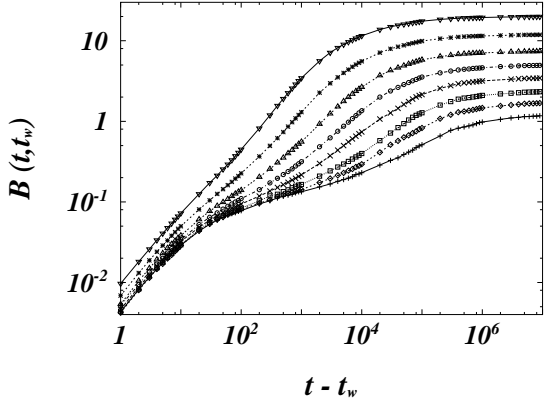,width=8cm}
\end{center}
\caption{ Mean squared displacement $B(t,t_{\rm w})$ for waiting times
	$t_{\rm w}=2^{10+k} (k=1, 2,...)$ and vibration amplitude
	$x=0.4$.}
	\label{fig.msd}
\end{figure}

\begin{figure}[h]
\begin{center}
\epsfig{file=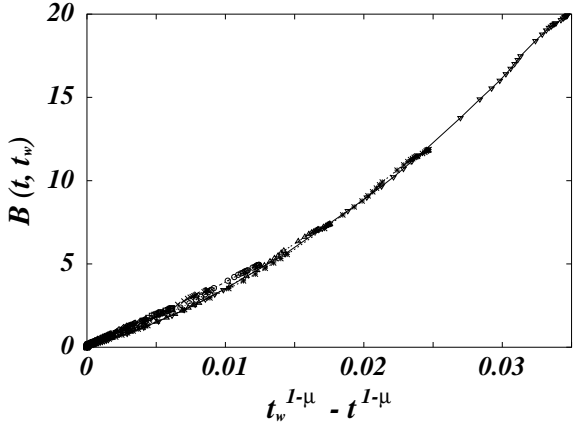,width=8cm}
\end{center}
\caption{ 
	Mean squared displacement $B(t,t_{\rm w})$ for several
	values of $t_{\rm w}$ and $x=0.4$ as a function of 
	$t_{\rm w}^{1-\mu}- t^{1-\mu}$.  The data collapse is 
	obtained for $\mu=1.48$.}  
	\label{fig.collapse}
\end{figure}

A similar super-aging behavior has been observed in the simulation of
the gravity-driven KA model~\cite{Sellitto01} where the agreement is
rather good if the vibration is not too low. For example, we show in
fig.~\ref{fig.msd} the mean square displacement for the case $x=0.4$
along with fig.~\ref{fig.collapse} where these curves were collapsed
onto a master curve following the above scaling.  The super-aging
exponent obtained from the data collapse is $\mu=1.48$ which is in very
good agreement with the theoretical prediction
$\mu=\phi/(\phi-1)\simeq 1.476$. For smaller vibrations, the exponents
become  smaller since the time accessible to the
measurement is probably not enough to reach the asymptotic regime
(where the approximation is valid), thus differing from the theoretical
value.

For a Vogel-Fulcher diffusion instead we find, to the leading order in
$t$ and $t_{\rm w}$, and for $z<z_{\rm 0}(\rhor)$:
\begin{equation}
	B_z(t,t_{\rm w}) \sim \log \left( \frac{t}{t_{\rm w}} \right)\,,
	\label{eq.bvogel}
\end{equation}
that is a simple aging scenario, whereas for the upper layers,
$z>z_{\rm 0}(\rhor)$, one obtains (to the leading order in $t$ and
$t_{\rm w}$) that the two-time scaling does not depend on $t/\tw$ but
rather
\begin{equation}
	B_z(t,t_{\rm w}) \sim {\rm e}^{-a \rhoc \tw } - 
	                      {\rm e}^{-a \rhoc t   } \,.
\end{equation}

\subsection{Triangle relation}

\begin{figure}[h]
\begin{center}
\epsfig{file=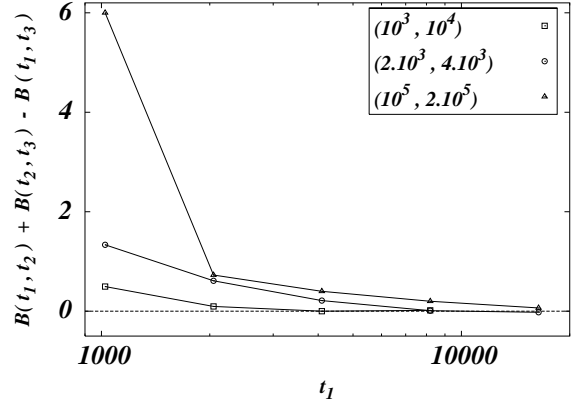,angle=270,width=8cm}
\end{center}
\caption{
	Check of the triangle relation for the mean squared
	displacement $B(t,t')$ for $x=0.1$. Different symbols 
	stand for different pairs $(t_2,t_3$) as indicated in
	the figure.}
	\label{fig.triangle}
\end{figure}

It is interesting to observe the different behavior of
Eqs.~\ref{eq.bvogel} and~\ref{B_a} at finite waiting times $t_{\rm
w}$.  In the case of simple aging, $\lim_{t \rightarrow \infty}
B(t,t_{\rm w}) = \infty$, i.e., a weak ergodicity
scenario~\cite{BoCuKuMe97}; while for super-aging, a finite limit is
obtained (which, however, vanishes as $t_{\rm w} \to \infty$).  The
manner in which time-translation invariance is violated is, however,
similar.  Indeed, if we consider the ``triangle relation'',
$B(t_1,t_3) = f \left[ B(t_1,t_2) ,\, B(t_2,t_3) \right]$, where the
times $t_1,\, t_2$, and $t_3$ are in increasing order, it is
straightforward to check that $f(x,y)=x+y$ in both Vogel-Fulcher and
power law cases, implying that displacements over non-overlapping time
intervals are statistically independent.  This feature does not hold
in the presence of activated aging for which $B(t,t_{\rm w}) \sim \log
t/\log t_{\rm w}$~\cite{BoCuKuMe97}.  In the gravity driven KA model
the triangle relation is not obeyed at short times but becomes
asymptotically valid at longer times as it can be seen in
fig.~\ref{fig.triangle},\footnote{We thank J. Kurchan for suggesting to
check the triangle relation.}.

\section{Dynamic heterogeneities}
\label{section.het}

In purely kinetic models, the absence of an increasing static
correlation lenght on approaching the dynamic arrest poses the
question of whether the increasing relaxation times can be related to
a diverging dynamic correlation lenght.  This has been associated to
the presence of dynamical heterogeneities in glasses(for a review
see~\cite{Sillescu99,Richert02,RiSo02} and references therein). If the
glass/dense granular analogy holds true then one would expect that the
role played by dynamical heterogeneities in slow granular compaction
should be similar to that observed in glassy dynamics.  However, the
role played by these structures and the associated lenghts, on the
dynamics of granular and colloidal systems, is yet to be understood.
Several measures for quantifying the spatial heterogeneities have been
introduced for kinetic models~\cite{RiSo02}. In particular, this issue
was recently investigated in the KA model~\cite{FrMuPa02} without
gravity using a fourth-order correlation function. Here we study the
role played by these structures in the non-zero gravity case.

\begin{figure}[h]
\begin{center}
\epsfig{file=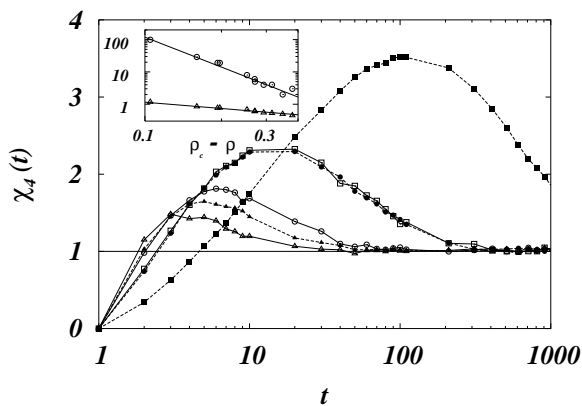,angle=270,width=8cm}
\end{center}
\caption{ Dynamical response, eq.~\ref{eq.het}, as a function of time
         (in MCS) for different vibrations: x=0.92 (filled symbols)
         and 0.94 (empty symbols).  Different symbols stand for
         different heights: $z=5, 10$ and 15 (square, circle,
         triangle, respectively).
         The line is the asymptotic $\chi_4=1$ behavior.  Notice in
         the figure the presence of two very close curves: they
         correspond to different vibrations and heights, but their
         density is the same within the numerical accuracy.  Inset:
         location (circles) and height (triangles) of the maximum of
         $\chi_4$. Both diverge as power laws with approximate
         exponents 3 and 0.6, respectively.}
\label{fig.het}
\end{figure}

In figure~\ref{fig.het} we plot the dynamical nonlinear response
\begin{equation}
\chi_4(z,t)=N \left( \left\langle q^2(z,t) \right\rangle - 
\left\langle q(z,t) \right\rangle^2 \right)
\label{eq.het}
\end{equation}
where $N$ is the number of involved sites in the computation,
$q(z,t)=C(z,t)/C(z,0)$ and
\begin{equation}
C(z,t) =\frac{1}{N}\sum_i n_i(t)n_i(0) - \rho(z,t)\rho(z,0)
\end{equation}
where $i$ runs over all sites in the $z$, $z-1$ and $z+1$ layers.
Consistently with the theoretical expectation~\cite{RiSo02}, the long
time limit of $\chi_4$ converges to unit,
$\chi_4(z,\infty)=1$~\footnote{We thank P. Sollich for discussions on
this point}. We also verified that this asymptotic behavior is valid
in absence of gravity at variance with the results of
Ref.~\cite{FrMuPa02}~\footnote{We thank R. Mulet for discussions on
this point}.  Analogously to what happens in the KA model without
gravity and in other glassy systems, the peak is shifted to higher
times and gets larger as the density increases (the lower is $z$, the
greater is the density).  In the inset of fig.~\ref{fig.het} we show
that both the position and height of the peak grow as power laws as
the density of the corresponding height approaches $\rhoc$.
Interestingly, $\chi_4$ only depends on the local density: for
example, in fig.~\ref{fig.het} is shown that two curves corresponding
to different $z$ and $x$, but having almost the same density, are
coincident within the numerical accuracy.

\section{Conclusions}
\label{section.conclusions}

We have investigated some aspects of slow granular dynamics inspired
by kinetic lattice-gas models~\cite{KoAn93,SeAr00}.  The key ingredient of
these models is a free-volume restriction implemented by a purely
kinetic constraint.  No interaction between the particles is assumed
beyond the hard core exclusion. The thermodynamics of the model is
completely trivial and all its interesting features are purely
dynamical. A macroscopic transport equation was written and studied,
allowing us to predict the specific location of a jamming transition
and to analyze the behavior of the system in its vicinity.

As noted in the earlier work~\cite{LeArSe01} and detailed here, the
time evolution of particle density in a gravity driven lattice gas is
completely controlled by the mobility of the corresponding
gravitationless system.  It was shown that for a power law mobility,
the bulk density relaxes as a power law.  On the other hand, when the
particle diffusion decreases exponentially accordingly to the
Vogel-Fulcher law, a logarithmically slow compaction is found.  Due to
the finite time window available, the data on granular systems from
the Monte Carlo simulation, as well as from experiments, is consistent
with both the power law and logarithmic relaxation.  Furthermore, the
issue of whether the asymptotic density is a monotonic function of the
vibration amplitude, can only be solved by performing simulations over
a much larger time window~\cite{Head00}.  For the maximum time
achieved here, the asymptotic packing density seems to depend on the
vibration strength in a non trivial way.

The similarities between granular and glassy systems have been
stressed many times in recent years. Here we extended this discussion
by showing that the behavior of dynamical heterogeneities is quite
similar to the ones present in systems without gravity. We showed that
for different vibrations the global non homogeneity induced by gravity
(that is, the density profile, different for each vibration) does not
affect the local character of these quantities that only depends on
the local density: the heterogeneities present in a horizontal layer
depends only on this layer density. This result could also be relevant
to the investigation of slow sedimentation of colloids.

Other signatures of glassy behavior are also present, like
aging, reversible-irreversible cycles and memory effects.
In particular, short and long-term memory effects are simpler 
than their glassy counterpart and have already been described in 
terms of the density profile properties. 
Irreversible branches
are obtained when cycling the vibration amplitude, approaching
a quasi-reversible branch when the rate is slow enough or
after cycling many times. 
We also remark the analogy between short-term memory effects
and the cycling experiments. In the decreasing-$x$ part of the cycles,
the density increases as for smaller $x$. This is at variance with
what one would expect from the long time behavior of the asymptotic
packing density, exactly in the same way as the system behaves in
short-term memory experiments. The difference is that in one case
the change in vibration  is discontinuous while in the other, it
is done at a small rate.

The analytical approach presented here has some limitations.  Among
the features seen in the simulation, which are not properly described
by the theory, are: existence of a dense layer between the bulk and
the interface and the initial state dependent oscillations at the
bottom of the sample.  These may be a direct consequence of the local
density approximation used for the mobility. A weighted density
approximation might be able to account for some of these features.
Moreover, since the equation is deterministic, fluctuation
dependent quantities ({\em e.g.}, $\chi_4(t)$) are not captured
by the formalism and noise has to be included~\cite{Stariolo97}.

Another interesting question concerns the role of friction. In
realistic systems, the energy injected by stirring, shearing, tapping
etc. is dissipated by the inelastic collisions among the particles,
creating a gradient of granular
temperature~\cite{WiHuHaPaAl00,WiHuPa01,YaHuCaMaWa02}.  Here we assume
that the system is in contact with a thermal bath, a situation which
is quite different.  In highly packed granular systems however the
steric hindrance is far more important than the energy dissipation, as
one is able to reproduce several features of compaction dynamics by
ignoring the specific mechanism of energy injection/dissipation.

Finally, we mention that slow relaxation in the KA model has also been
analyzed in terms of the dynamically available volume, expressed by
holes density $\nu$~\cite{LaReMcGrTaDa02}.  A hole is defined as an
empty site where a neighbouring particle satisfies the kinetic
constraints and may jump in. It has been recently suggested that
dynamical arrest in different systems, such as glasses, colloids,
gels, compressed emulsion and granulars, has a universal character
when described in terms of $\nu$. In particular, as one approaches the
transition, the diffusion coefficient goes to zero as a power-law
$D=(\nu-\nu_0)^{\gamma}$, where $\nu_0$ is the residual density of
rattlers that do not contribute to the macroscopic diffusion, and
$\gamma$ is an universal exponent that assume the value $\gamma=2$
irrispective of the system considered~\cite{LaReMcGrTaDa02}. It would
be interesting to investigate to what extent the presence of gravity,
and consequently of a heterogeneous density profile might modify the
above results.

In conclusion, nonlinear diffusion equation and kinetically
constrained models, in spite of their simplicity, seem to capture the
main dynamical features of dense granular materials and provide a
natural framework in which slow relaxation phenomena in granular and
glassy matter can be easily understood.

\begin{acknowledgments}
This work was supported in part by the Brazilian agencies
CNPq and FAPERGS.  JJA thanks the Abdus Salam ICTP for hospitality
where part of this work was done.
\end{acknowledgments}



\end{document}